\documentclass[
reprint,
superscriptaddress,
amsmath,amssymb,
aps,
prl,
floatfix,
]{revtex4-1}

\usepackage{graphicx}
\usepackage{dcolumn}
\usepackage{bm}
\usepackage{hyperref}
\usepackage{mathtools}
\usepackage{ulem}
\usepackage{float}
\usepackage[usenames, dvipsnames]{xcolor}

\DeclarePairedDelimiterX\braket[2]{\langle}{\rangle}{#1 \delimsize\vert #2}

\newcommand{\vdot}{V_\mathrm{dot}}
\newcommand{\vref}{V_\mathrm{Ref}}
\newcommand{\vb}{V_\mathrm{b}}

\graphicspath{{figures/}}

\begin{document}


\title{Quantum-Dot Parity Effects in Trivial and Topological Josephson Junctions}

\author{D.~Razmadze}
 $ \thanks{These authors contributed equally to this work}$
\affiliation{Center for Quantum Devices, Niels Bohr Institute, University of Copenhagen, 2100 Copenhagen, Denmark}
\affiliation{Microsoft Quantum Lab--Copenhagen, 2100 Copenhagen, Denmark}

\author{E.~C.~T.~O'Farrell}
\affiliation{Center for Quantum Devices, Niels Bohr Institute, University of Copenhagen, 2100 Copenhagen, Denmark}
\affiliation{Microsoft Quantum Lab--Copenhagen, 2100 Copenhagen, Denmark}

\author{P.~Krogstrup}
\affiliation{Center for Quantum Devices, Niels Bohr Institute, University of Copenhagen, 2100 Copenhagen, Denmark}
\affiliation{Microsoft Quantum Materials Lab--Copenhagen, 2800 Kongens Lyngby, Denmark}

\author{C.~M.~Marcus}
\affiliation{Center for Quantum Devices, Niels Bohr Institute, University of Copenhagen, 2100 Copenhagen, Denmark}
\affiliation{Microsoft Quantum Lab--Copenhagen, 2100 Copenhagen, Denmark}


\begin{abstract}
An odd-occupied quantum dot in a Josephson junction can flip transmission phase, creating a $\pi$-junction. When the junction couples topological superconductors, no phase flip is expected. We investigate this and related effects in a full-shell hybrid interferometer, using gate voltage to control dot-junction parity and axial magnetic flux to control the transition from trivial to topological superconductivity. Enhanced zero-bias conductance and critical current for odd parity in the topological phase reflects hybridization of the confined spin with zero-energy modes in the leads. 
\end{abstract}

\maketitle

The development of topologically protected qubits \cite{Alicea2011,Aasen} for quantum computing \cite{Kitaev2003,RevModPhys.80.1083} benefits from fundamental investigations that examine signatures of topological superconductivity in various device geometries. These serve both to test theoretical models and solidify the interpretation of experiments \cite{Cayao2016,PhysRevB.91.024514}. A fruitful system for exploring topological states is based on semiconductor nanowires with strong spin-orbit coupling in contact with a metallic superconductor \cite{Oreg2010,Lutchyn2010,Luthcyn2018NatMat}. Recently, semiconductor nanowires with a fully surrounding superconducting shell were found to offer a convenient means of tuning into the topological phase using applied axial magnetic flux \cite{SoleScience}. In this system, the destructive Little-Parks effect \cite{little_parks}, with associated winding of superconducting phase around the shell, induces a topological phase in the semiconductor core.

Here, we investigate Josephson junctions realized in full-shell InAs/Al nanowires, focussing on parity effects of a gate-controlled quantum dot in the junction. We investigate even and odd occupancies of the dot for the zeroth and first lobes of the reentrant Little-Parks structure in the leads. The hybrid nanowire containing the dot-junction is embedded in a superconducting interferometer, allowing phase across the dot-junction to be measured relative to a reference arm containing a second gate-controlled junction. Depleting the reference junction {\it in situ} with a gate voltage allowed the dot-junction to be measured in isolation, revealing related parity-dependent features in conductance.

Differential conductance of the isolated dot-junction as a function of applied voltage bias showed a strong zero-bias peak throughout the first lobe only for an odd-occupied dot-junction, reminiscent of Kondo-enhanced zero-bias conductance peaks \cite{Buitelaar2002, Choi2004, Lee2012} seen for odd-occupied dots with superconducting leads \cite{Buitelaar2002, Choi2004, Grove2007, Sand2007, Eichler2007, Buizert2007, Karrasch2008, Eichler2009, Kanai2010, Maurand2012, Lee2012, Chang2013_2, Zitko2015, Meden2019, Estrada2019, Kadlecova2019}.  When the dot-junction had even occupancy, the zeroth and first superconducting lobes showed comparable conductance at all biases. Opening the reference junction, we observed a 0-$\pi$ transition as a function of dot occupancy in the zero lobe, consistent with previous studies \cite{VanDam2006, Jorgensen2007, Maurand2012, Delagrange2016}, while in the first lobe, the 0-$\pi$ transition was absent, consistent with recent predictions that hybridization of the dot spin with zero-energy states in the topological-superconducting leads favors a conventional current-phase relation, that is, a 0-junction \cite{Gao2015, Camjayi2017, Schrade2018, Awoga2019, Schulenborg2020}. We note that hybridization of odd-occupied dot spin with discrete zero-energy states in the leads is distinct from Kondo hybridization \cite{Cheng2014}, the latter also favoring a 0-junction, as discussed below \cite{Choi2004, Sellier2005, Maurand2012, Allub2015, Meden2019, Kadlecova2019}. 

\begin{figure}[t]
	\includegraphics[width=0.45 \textwidth]{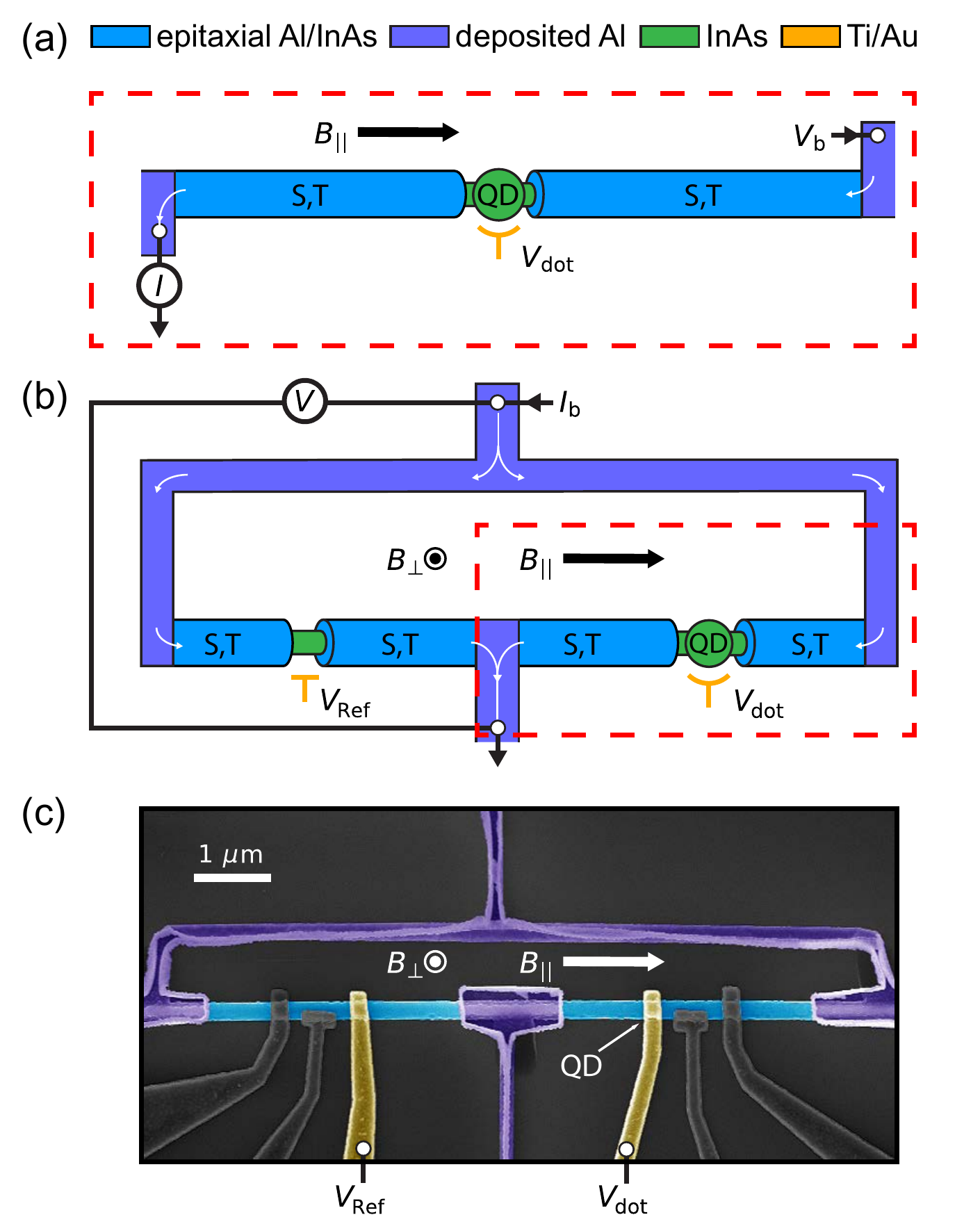}
	\caption{(a) Schematic of a dot-junction made from an InAs nanowire (green) containing a quantum dot (QD) with coupling and occupancy controlled by voltage $\vdot$. A voltage bias, $V_{\mathrm b}$, with a small AC component was applied across the single dot-junction and the current, $I$, measured. Thin Al leads (purple) remain superconducting with applied axial magnetic field, $B_{\parallel}$. The lobe structure in the destructive Little-Parks regime accesses trivial (S) or topological (T) superconductivity in the leads \cite{SoleScience}.  
	(b) The dot-junction was embedded in an interferometer with a reference junction controlled by gate voltage, $V_{\mathrm{Ref}}$.  Current bias, $I_{\mathrm b}$, with small AC component was applied and voltage, $V$, measured. Perpendicular magnetic field, $B_{\perp}$, controlled phase around the interferometer.  
	(c) False-color micrograph of a measured device, showing a loop of thin Al deposited on ramps to contact the full-shell wire. Uncolored gates were set to $+2\,$V.
}
	\label{fig:1}
\end{figure}

\begin{figure*}
	\includegraphics[width=1 \textwidth]{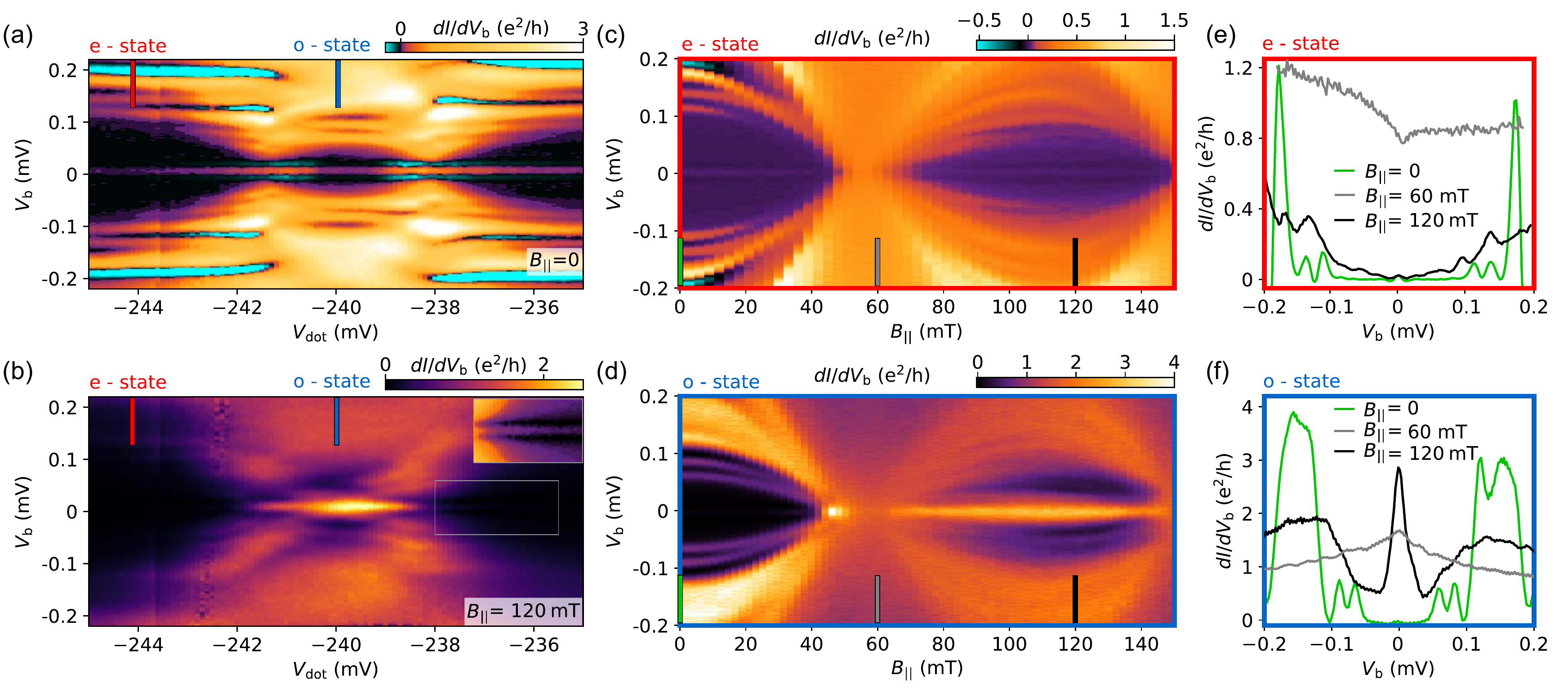}
	\caption{ Bias spectroscopy of the isolated dot-junction, with reference arm closed. (a)  Differential conductance, $dI/dV_{\mathrm b}$, as a function of dot-junction gate voltage, $\vdot$, and DC bias, $V_{\mathrm b}$, across the junction, in the zero lobe ($B_{\parallel}=0$). Sweeping $\vdot$ changes dot occupancy from even ({\it e}-state) to odd ({\it o}-state) to even. A uniform conductance (supercurrent) peak at $V_{\mathrm b}=0$  is visible throughout the range of $\vdot$. Negative differential conductance features (green) are visible in the {\it e}-state. Red (blue) marks indicate location of $e$($o$)-state cuts in (c(d)).   (b) Same as (a) except in the first lobe ($B_{\parallel}=120$~mT). A strong enhancement of the zero-bias conductance peak occurs in the odd occupied state. (c) Lobe structure in bias spectroscopy as a function $B_{\parallel}$ for {\it e}-state. Green, gray, and black marks indicate cuts in (e). (d) Same as (c) for {\it o}-state, with green, gray, and black marks indicating cuts in (f). Enhanced zero-bias conductance persists through the first lobe, and does not split with increasing magnetic field. We interpret the isolated zero-bias peak at $B_{\parallel}\sim46$~mT as a Kondo enhancement (see text). (e) Cuts from (c) in the {\it e}-state showing small supercurrent peaks and several subgap resonances in both the zeroth (green) and first (black) lobes, with a broad zero-bias dip in the destructive regime (gray). (f) Cuts from (d) in the {\it o}-state, showing a large zero-bias conductance peak in the first lobe and a broad zero-bias peak in the destructive regime (gray). }
	\label{fig:2}
\end{figure*}

Supercurrent through a conventional Josephson junction is given by $I = I_{\mathrm{c}}\,{\rm sin\,} (\varphi)$, where $I_{c}$ is the critical current and $\varphi$ is the phase difference of the superconducting order parameter across the junction. In few-channel junctions, including the semiconductor junctions investigated here, higher harmonics of $I(\varphi)$ are also present, but the basic periodicity, $I(\varphi) = I(\varphi + 2\pi)$, and symmetry, $I(\varphi) = -I(-\varphi)$, remain \cite{Spanton2017}. Symmetry upon reversing phase can be lifted for particular arrangements of magnetic and spin-orbit fields \cite{Zazunov2009, Szombati2016}, including a predicted  supercurrent at zero phase across a dot-junction near a single-triplet anticrossing with topological leads \cite{Schrade2017}. Lifting of $2\pi$ periodicity by Majorana coupling \cite{FuKane2009, Veldhorst2012} is not observed.

As discussed in recent proposals \cite{Gao2015, Camjayi2017, Schrade2018, Awoga2019, Schulenborg2020}, the transmission phase through a quantum dot embedded in a Josephson junction---a well-studied system, see experimental \cite{Franceschi2010} and theoretical \cite{Martin2011, Meden2019} reviews---provides a means of investigating topological superconductivity. Coulomb energy of the dot-junction suppresses Cooper-pair tunneling, relying on spin-dependent cotunneling processes, which in turn depend on dot  occupancy \cite{Glazman1989, Spivak1991, Rozhkov, Vecino2003, VanDam2006, Delagrange2016, Kadlecova2019}. In its simplest form, for even dot parity ({\it e}-state), the phase across the junction matches the conventional current-phase relation, while for odd parity ({\it o}-state), supercurrent typically involves a sign reversal, $I = I_{\mathrm{c}}\,{\rm sin}\,(\varphi+\pi)  = -I_{\mathrm{c}}\,{\rm sin}\,(\varphi)$, resulting in a supercurrent reversal, or $\pi$-junction.

InAs nanowires with $\sim 130$~nm diameter were grown by molecular beam epitaxy using the vapor-liquid-solid method, followed by {\it in-situ} growth of a $\sim 30$~nm epitaxial Al shell fully surrounding the semiconductor core \cite{Krogstrup2015}. After placing the nanowires on a Si/SiO$_{2}\,$ substrate, polymer ramps were patterned  to connect a loop and leads made of 25 nm of deposited Al, as shown in Fig.~\ref{fig:1}(c). The thin Al ensured that superconductivity was maintained in moderate fields along the nanowire axis. An insulating layer of HfO$_{2}$~(7~nm) was then deposited, followed by patterned Ti/Au top-gates used to control electron density in regions where the Al was removed by wet etching. An electron micrograph of one of the devices is shown in Fig.~\ref{fig:1}(c), with false-colored active regions and uncolored gates set to $+2$~V. All wire segments exceed 1~$\mu$m, several times the Majorana localization length, $\xi\sim180\,$nm \cite{SoleScience}. 

Measurements were carried out in a dilution refrigerator with a base electron temperature of $\sim 50$~mK using conventional lock-in techniques in both voltage-bias and current-bias configurations. A vector magnet provided independent control of magnetic field along the wire axis, $B_{\parallel}$, and a small transverse field, $B_\perp$, used to apply flux to the interferometer loop. A total of ten devices were cooled. Three devices were reasonably stable and showed qualitatively similar behavior. One of those is presented in the main text and the other two in the Supplementary Material (SM) Figs.~S1-S5. Among the others, three were nonconducting or did not show a supercurrent, two showed excessive noise and did not have a controllable dot in the junction, one did not show a $\pi$-junction in the zeroth lobe, and one appeared non-topological, without zero-bias feature in the first lobe and a $\pi$-junction in both lobes.  

With the reference arm closed by setting $\vref=-2\,$~V, the  dot-junction was measured in a voltage-bias configuration, applying ac+dc voltage $V_{\rm b}$ ($2\, \mu$V ac excitation),  and separately measuring ac and dc current, as shown in Fig.~\ref{fig:1}(a).  At negative $\vdot$, approaching depletion, sharp resonances in tunneling conductance, $dI/dV_{\rm b}$, were observed, indicating that a Coulomb blockaded quantum dot has formed in the junction. Note that $\vdot$ controlled both the dot-junction occupancy and, on larger voltage scales, the coupling to the leads. Tunneling spectra at $B_\parallel=0$, across a range of $\vdot$ spanning two {\it e}-states and one {\it o}-state are shown in Fig.~\ref{fig:2}(a). A narrow supercurrent feature at zero bias can be seen throughout the sweep with two enhancements at the charge transition points, corresponding to Coulomb blockade resonances. We note regions of negative differential conductance in the zeroth lobe [green stripes in Fig.~\ref{fig:2}(a)] at low bias near the charge transitions, and at higher bias in the {\it e}-states.  These features presumably reflect the opening of weakly coupled transport channels of the dot that blockade transport once populated \cite{Hekking1993}. Their prevalence in the {\it e}-state indicates spin-dependent excited states for even occupancy. 

Applying $B_{\parallel}$ reveals the lobe structure of destructive superconductivity, with suppressed superconductivity around $B_\parallel=50-60\,$~mT and a first lobe centered around $B_\parallel=120\,$~mT, corresponding to one quantum of applied flux and one twist of superconducting phase round the shell circumference. Figures~\ref{fig:2}(a,b) reveal a striking difference in bias spectra of the lobes. In particular, the first lobe [Fig.~\ref{fig:2}(b)] showed strongly enhanced zero-bias conductance in the {\it o}-state but not in the {\it e}-state, while spectra in the zeroth lobe showed similar conductance for both occupancies [Fig.~\ref{fig:2}(a)].  Bias spectra as a function of $B_{\parallel}$ in Figs.~\ref{fig:2}(c,d) show a complementary view: In the zeroth lobe, {\it o}-state and {\it e}-state spectra are comparable, while throughout the first lobe the zero-bias conductance is strongly enhanced only for the {\it o}-state, with enhancement roughly tracking the size of the topological gap. Cuts in Figs.~\ref{fig:2}(e,f) show a large zero-bias conductance peak in the first lobe for the {\it o}-state, with  $12\, \mu$V half-width at half maximum. Cuts along zero bias as a function of $B_{\parallel}$ are shown in SM Fig.~S6. We note that the zero-bias peak in the {\it o}-state in the first lobe does not appear to split with increasing $B_{\parallel}$. For a conventional Kondo peak in conductance, for instance arising from a soft gap in the first lobe \cite{Lee2012}, the peak would be split by $2g\mu_{B}B_{\parallel} > 50\mu$eV in the first lobe, which would be visible. 

We note in Fig.~\ref{fig:2}(d) a small, bright zero-bias peak at the closing of the zeroth lobe, $B_{\parallel}\sim46$~mT. This small feature does not persist further into the zeroth lobe or into the destructive regime, where instead a broad zero-bias peak can be seen [gray cut in Fig.~\ref{fig:2}(f)], while in the {\it e}-state, the destructive regime had a zero-bias dip [gray cut in Fig.~\ref{fig:2}(e)]. The bright peak at $B_{\parallel}\sim46$~mT is more easily seen in the cut in SM Fig.~S6. We interpret the narrow peak at $B_{\parallel}\sim46$~mT as Kondo-enhanced conductance in the superconducting regime \cite{Buitelaar2002, Choi2004, Lee2012}. From the ratio of superconducting to  normal conductance, $G_{S}/G_{N}\sim 2$, [from Figs.~\ref{fig:2}(d) and S6] we infer a rough ratio of Kondo temperature to gap, $T_{K}/\Delta \sim 2$ \cite{Buitelaar2002, Choi2004}. Within this interpretation, the width of the peak and its appearance only at the closing of the zero lobe suggests a low $T_{K}$, of order $10\,\mu$eV. The zero-bias peak in the {\it o}-state destructive regime presumably reflects normal-state Kondo enhancement.

\begin{figure}[t!]
	\includegraphics[width=0.5\textwidth]{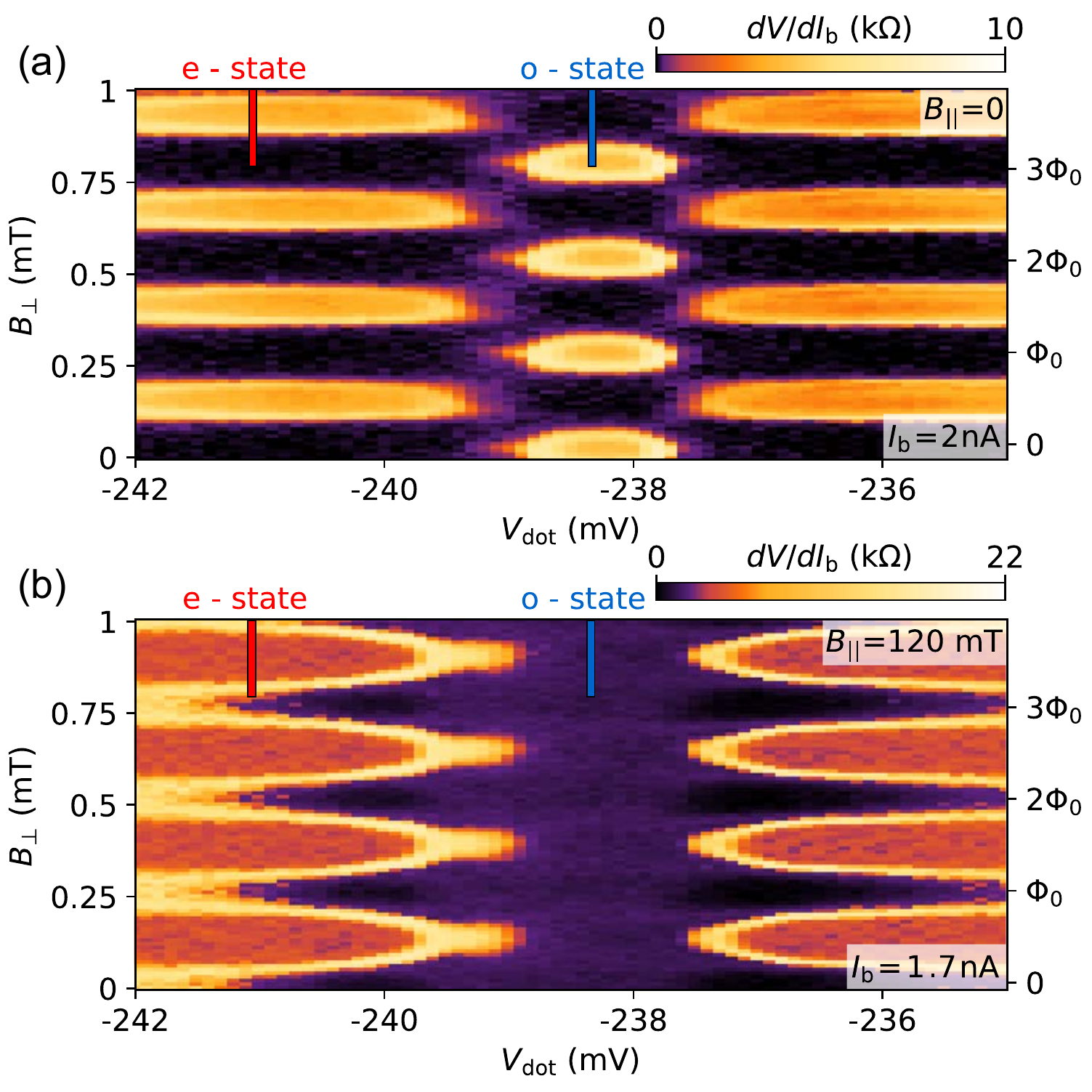} 
	\caption{
		Differential resistance, $dV/dI_{\mathrm b}$, of the interferometer as a function gate voltage, $\vdot$, controlling dot occupancy, and $B_\perp$, controlling flux through the interferometer. Current bias, $I_{\mathrm{b}}$, was set to periodically exceed the total switching current of the interferometer.  (a) The zeroth lobe ($B_{\parallel}=0$) with $I_{\mathrm{b}}=2$~nA, showed a $\pi$ phase shift in the {\it o}-state relative to the {\it e}-state, indicating a $\pi$-junction.
		(b)  Same as (a) except in the first lobe ($B_\parallel=120$~mT) with $I_{\mathrm{b}}=1.7$~nA, showing no phase shift as a function of $\vdot$.}
	\label{fig:3}
\end{figure}

Opening the reference arm by setting $\vref = 0$~V connected the interferometer loop, yielding a switching current of 2~nA in the reference junction, compared to $\sim 1$~nA in the dot-junction. In the configuration of Fig.~\ref{fig:1}(b), whenever the current bias $I_{\mathrm{b}}$ exceeded the total switching current of the interferometer a finite differential resistance, $dV/dI_{\mathrm{b}}$, appeared across the interferometer. Figure~\ref{fig:3}(a) shows $dV/dI_{\mathrm{b}}$ for dc current bias $I_{\mathrm{b}}=2$~nA (with ac excitation 0.2~nA) as a function of $\vdot$ and $B_{\perp}$ in the zeroth lobe, with $B_{\parallel} =0$. To avoid hysteretic effects, $I_\mathrm{b}$ was briefly set to zero then reset to 2~nA for each data point (pixel) in the two-dimensional plot. Figure~\ref{fig:3}(a) shows the periodic dependence of the zero-resistance state with magnetic flux through the interferometer, consistent with $\Delta B_\perp A = \Phi_{0} = h/2e\,$, where $A$ is the area of the interferometer [see micrograph in Fig.~\ref{fig:1}(c)]. As $\vdot$ was swept from the {\it e}-state to the {\it o}-state, the phase of oscillation with $B_{\perp}$ shifted by $\Phi_{0}$/2, indicating that the dot-junction is a $\pi$-junction in the {\it o}-state relative to the {\it e}-state.

  \begin{figure}[tbh]
  	\includegraphics[width=0.5\textwidth]{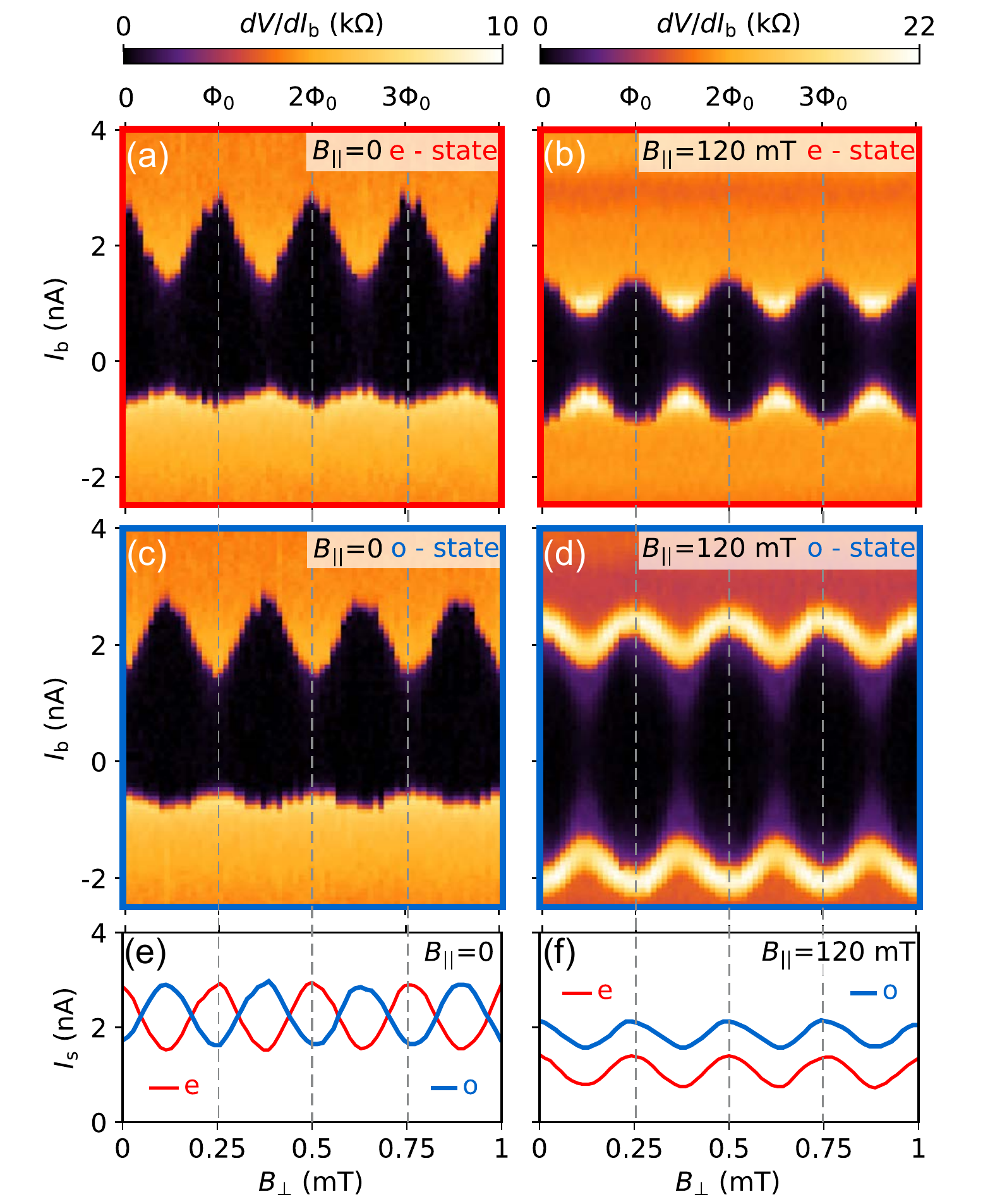} 
	  \caption{Differential resistance, $dV/dI_{\mathrm b}$, of the interferometer as a function of bias current, $I_\mathrm{b}$, and perpendicular magnetic field, $B_\perp$, (a,c) in the zeroth lobe, along cuts through the $e$($o$)-state [red(blue) marks in Fig.~\ref{fig:3}(a)], showing relative $\pi$ phase shift, and (b,d) in the first lobe, along cuts through the $e$($o$)-state [red(blue) marks in Fig.~\ref{fig:3}(b)], showing absence of phase shift. Note in (a,c) that switching currents exceed retrapping currents. (b,d) In the first lobe ($B_{\parallel}=120$~mT), switching and retrapping currents are comparable. (e) Relative phase shift of $\pi$ between  {\it e}-state (red) and {\it o}-state (blue) in the zeroth lobe. (f) no phase shift between the {\it e}-state (red) and {\it o}-state (blue) in the first lobe ($B_{\parallel}=0$), where both phases align with the {\it e}-state (red) in the zeroth lobe. Critical current in the {\it o}-state (blue) exceeds the {\it e}-state (red) in the first lobe ($B_{\parallel}=120$~mT).}
	  
	\label{fig:4}
  \end{figure}

Figure~\ref{fig:3}(b) shows a similar plot of $dV/dI_{\mathrm{b}}$ as a function of  $B_\perp$ and $V_{\mathrm{dot}}$, now in the first lobe, taken at $B_{\parallel} = 120\,$mT, demonstrating the {\it absence} of a $\pi$ phase shift for relative occupancies. The absence of $\pi$-junction behavior for topological dot-junctions is consistent with theoretical predictions \cite{Schrade2018, Awoga2019, Schulenborg2020}. Oscillations in $dV/dI_{\mathrm{b}}$ are less visible in the {\it o}-state in the first lobe [Fig.~\ref{fig:3}(b)] compared to the zeroth lobe [Fig.~\ref{fig:3}(a)]. This is because the switching current of the dot-junction was considerably larger in the {\it o}-state in first lobe, so that the total switching current of the interferometer barely exceeds the fixed bias $I_{\mathrm{b}}=2$~nA. This is more clearly seen by measuring $dV/dI_{\mathrm{b}}$ as function of a swept $I_{\mathrm{b}}$, as shown in Fig.~\ref{fig:4}. Differential resistance $dV/dI_{\mathrm{b}}$ of the interferometer along cuts through the {\it e}-state and {\it o}-state of the dot-junction showed oscillatory patterns of switching and retrapping currents with applied flux, noting that $I_{\mathrm{b}}$ was stepped from negative to positive. Retrapping occurs at negative values of $I_{\mathrm{b}}$. Similar data for the other devices are shown in SM Figs.~S4 and S6. Phase plots at other fields for device 1 are shown in SM Fig.~S7.

We draw attention to several features in Fig.~\ref{fig:4}: (i) There is a $\pi$ phase shift between panels (a) and (c), indicating that in the zeroth lobe, the {\it o}-state forms a $\pi$-junction relative to the {\it e}-state.  (ii) There is no $\pi$ phase shift between panels (b) and (d), indicating that in the first lobe there is no relative $\pi$-junction upon changing dot occupancy. We do not observe a nontrivial phase shift in (d), noting that $B_{\parallel}$ is probably too induce a single-triplet crossing \cite{Schrade2017}. (iii) The absolute phase of oscillations as a function of $B_{\perp}$ is the same in all four panels, with only the {\it o}-state in the zeroth lobe shifted by $\pi$ [panel (c)]. We note that phase was not corrected for a given $B_{\parallel}$. (iv) Retrapping currents in the zeroth lobe [panels (a,c)] are considerably smaller than switching currents, a consequence of underdamping and junction heating in the resistive state. Retrapping and switching currents are roughly the same in the first lobe, presumably due to subgap modes that both dampen junction dynamics and cool the junction through the leads. While these features warrant further study, we take this as possible indirect evidence for increased junction damping and thermal conductivity in the first lobe. (v) Finally, we observe that switching and retrapping currents in the {\it e}-state and the {\it o}-state are comparable in the zeroth lobe [panels (a,c,e)] whereas in the first lobe, switching currents are larger in the {\it o}-state than in the {\it e}-state [see panels (b,d,f)], as anticipated form the isolated dot-junction data in Fig.~\ref{fig:2}, where the {\it o}-state showed enhanced conductance compared to the {\it e}-state in the first lobe.

Finally, we revisit the small feature at $B_{\parallel}\sim46$~mT, interpreted above as Kondo-enhanced conductance at the closing of the zeroth lobe, now in the interferometer configuration. We observed enhanced critical current, $I_{c}\sim 1$~nA, and 0-junction behavior at that location, contrasting the $\pi$-junction behavior inside the zeroth lobe, as shown in SM Fig.~S8. The enhanced critical current is consistent with an estimate for an overdamped junction, $G_{S}/G_{N}\sim{\rm exp}(\hbar I_{c}/ek_{B}T)$ \cite{Choi2004}.  Taking $G_{S}/G_{N}\sim 2$ and temperature $T\sim 50$~mK yields $I_{c}\sim 1$~nA, close to the measured value. 


We thank A.~Akhmerov, K.~Flensberg, L.~Fu, L.~Glazman, J.~Paaske, C.~Schrade, J.~Schulenborg, G.~Steffensen, and S.~Vaitiek\.{e}nas for valuable discussions, and S.~Upadhyay for help with fabrication. Research is supported by Microsoft, the Danish National Research Foundation, and the European Research Commission, grant 716655.

\bibliography{references1}

\renewcommand{\thefigure}{S\arabic{figure}}
\setcounter{figure}{0}

\section*{Supplementary Information for \\"Quantum-Dot Parity Effects in Trivial and Topological Josephson Junctions''}

\section{Other devices}

\subsection{Device description}

\begin{figure}[t]
	\includegraphics[width=0.45\textwidth]{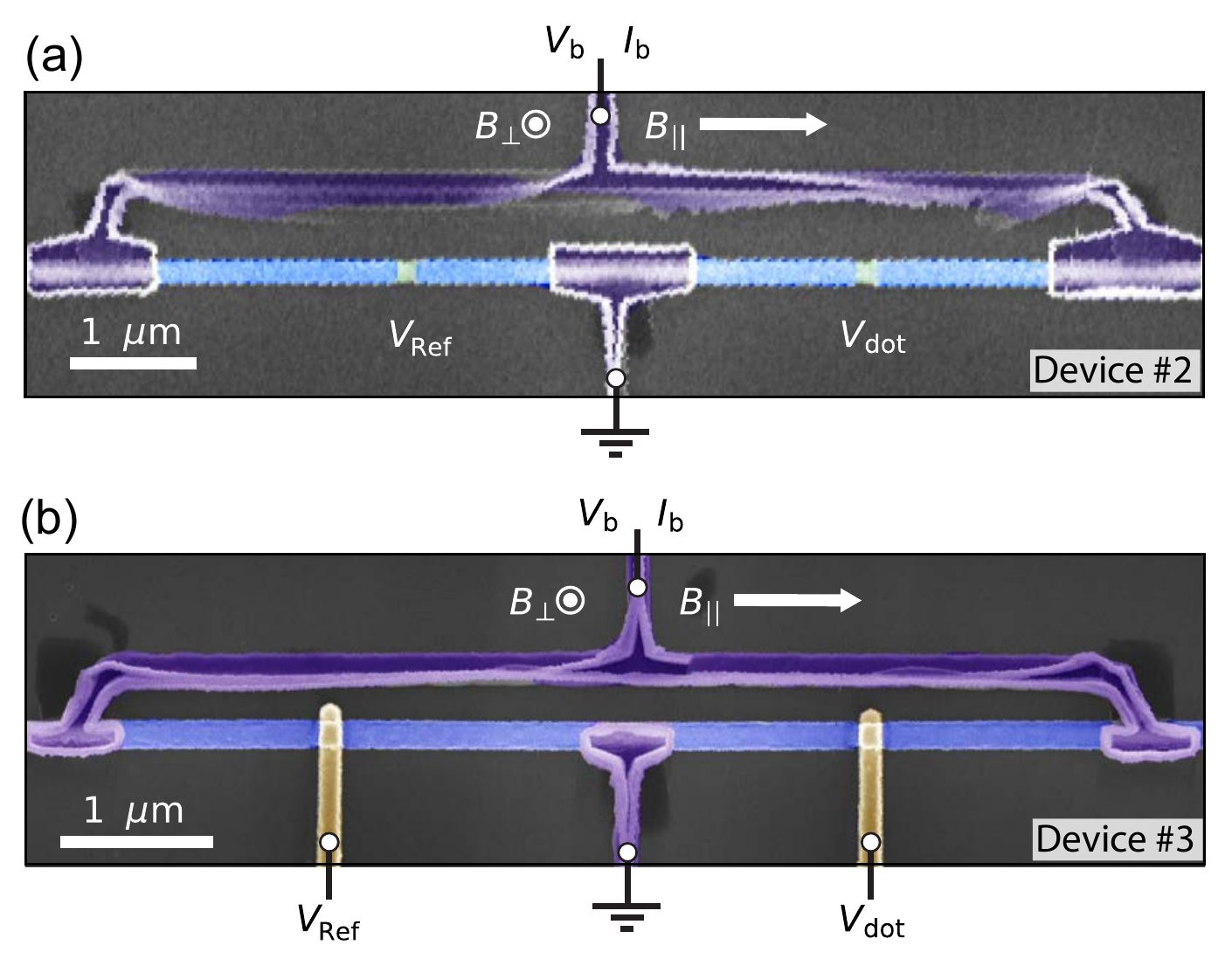}
	\caption{False-color scanning electron micrographs of other measured devices. (a) Device 2 with two controllable junctions $\vref$ and $\vdot$. Micrograph was taken before gate deposition, allowing regions of etched Al shell to be seen. (b) Device 3 micrograph, taken after fabrication of the gates, controlled by $\vref$ and $\vdot$. For both devices axial $B_\parallel$ and out-of-plane $B_\parallel$ magnetic fields are independently controlled.}
	\label{fig:S1}
\end{figure}

We present measurements of two other devices, denoted devices 2 and 3. The device in the main text is device 1. Micrographs of devices 2 and 3 are shown in Fig.~\ref{fig:S1}(a,b). The micrograph of device 2 was taken before final gate deposition so that the bare etched junctions are visible. The active elements used in this experiment were the same for all devices. In contrast to device 1, however, devices 2 and 3 do not have additional junctions (uncolored gates in Fig. 1 of the main text) that were  open during measurements. All devices were fabricated using the procedure described in the main text, using nanowires from the same growth batch with $\sim\,130\,$nm InAs diameter core and a 30~nm Al-shell. 

The axial magnetic field, $B_\parallel$ was used to control flux winding in the Al-shell. Perpendicular magnetic field, $B_\perp$ was used for phase measurements. $\vdot$ was used to control the occupancy of a quantum dot formed close to depletion in the dot-junction. We first describe the results of voltage-bias measurements, and then current-bias measurements in these devices.

\subsection{Voltage-bias measurements}

\begin{figure*}[!ht]
	\includegraphics[width=0.9\textwidth]{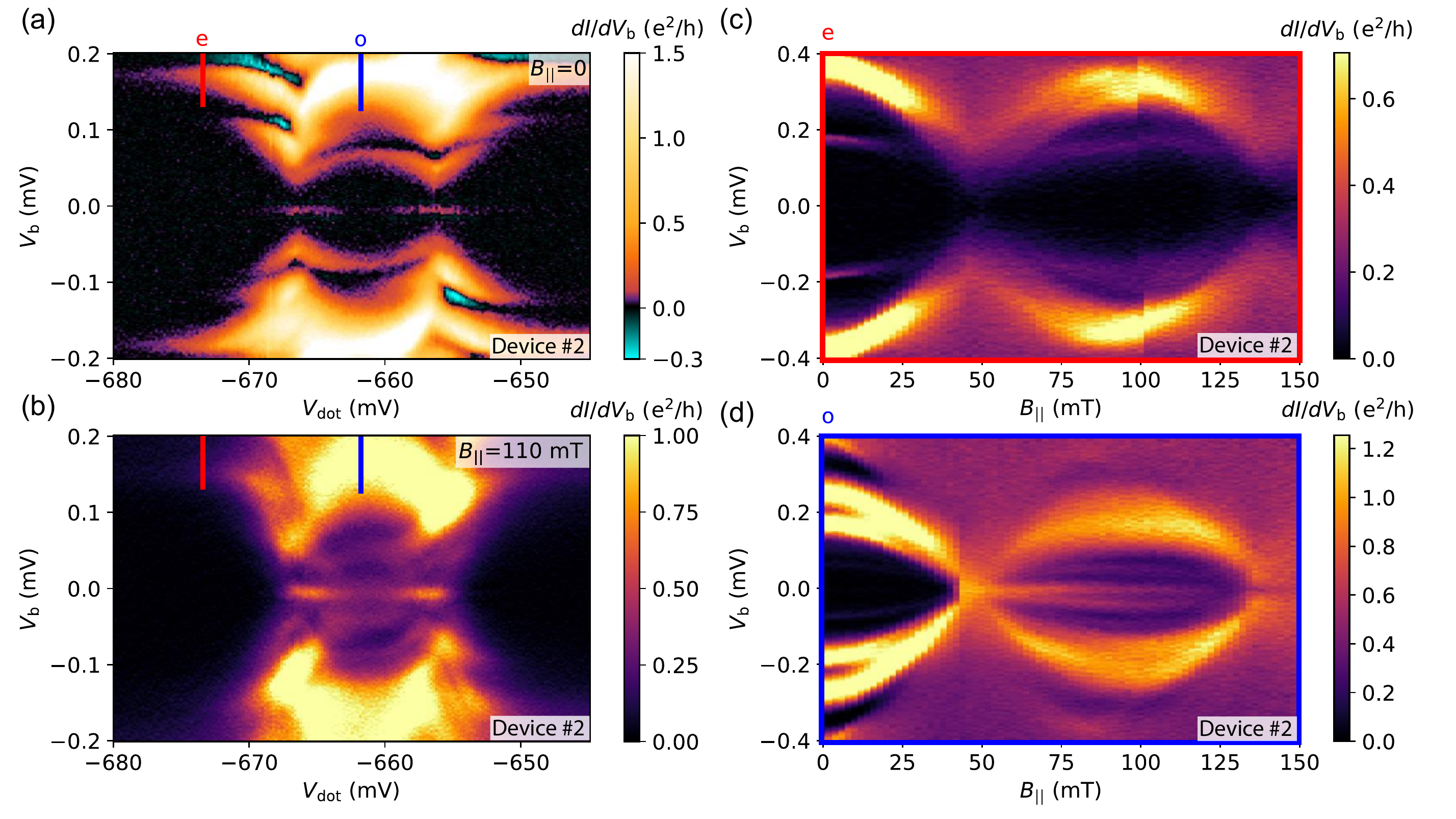}
	\caption{Voltage-bias measurement of isolated dot-junction of device 2 with reference arm closed. (a) Differential conductance $dI/dV_{\mathrm b}$ as a function of voltage-bias, $V_\mathrm{b}$, and $V_\mathrm{dot}$ at $B_\parallel=0$. Even ($e$) and odd ($o$) dot occupancies are indicated with red and blue markers. (b) The same voltage range of $\vdot$ gate indicating the same quantum-dot with two charge occupancies $(e,\,o)$ at $B_\parallel=110$mT. (c) Differential resistance measurement of axial magnetic field, $B_\parallel$, evolution as a function of voltage-bias in the even dot occupancy (d) Similar measurement as panel (c) for the $o$-state Coulomb valley.}
	\label{fig:device2_voltage}
\end{figure*}

Voltage-bias spectroscopy for both devices 2 and 3 on the dot-junction was performed by closing the reference arm of the interferometer by setting $\vref=-2\,$V.

Figure~\ref{fig:device2_voltage} shows voltage-bias measurements of the differential conductance for device 2. First, in the zeroth lobe ($B_\parallel=0$), Fig.~\ref{fig:device2_voltage}(a) shows differential conductance, $dI/dV_{\mathrm b}$, as a function of the DC bias, $V_\mathrm{b}$, and $V_\mathrm{dot}$. Two enhancements of the zero-bias conductance were observed at the charge transition points of a quantum dot formed in the junction. Figure~\ref{fig:device2_voltage}(b) shows conductance in the same range of $\vdot$ in the first lobe at $B_\parallel=110\,$mT; a significant enhancement of the zero-bias conductance peak was observed in the Coulomb valley, we assigned this valley as the dot $o$-state based current-bias measurements presented below. Figure~\ref{fig:device2_voltage}(c) shows conductance as a function of $B_\parallel$ in the $e$-state, demonstrating the closing of the superconducting gap by the Little-Parks effect. In contrast to the $e$-state, the $o$-state (Fig.~\ref{fig:device2_voltage}(d)) showed a strong enhancement of the zero-bias conductance in the first lobe, in a similar manner to device 1 in the main text.

\begin{figure*}[!ht]
	\includegraphics[width=0.9\textwidth]{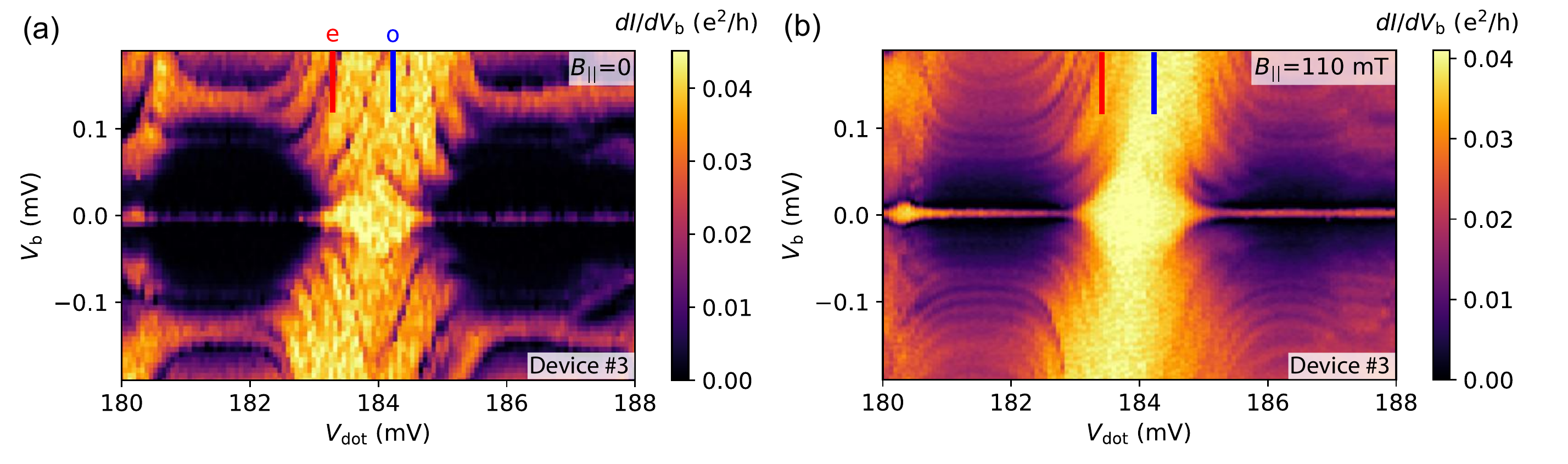}
	\caption{Voltage-bias measurement of isolated dot-junction in Device 3 with reference arm closed. (a) Differential conductance $dI/dV_{\mathrm b}$ as a function of voltage-bias $\vb$ and $vdot$, red and blue markers indicate even and odd dot occupancies, respectively. (b) Similar as panel (a) in the first lobe at $B_\parallel=110$mT.}
	\label{fig:device3_voltage}
\end{figure*}

Figure~\ref{fig:device3_voltage} shows voltage-bias measurements for device 3. Charge transitions are observed at $\vdot\sim184\,$mV in both the zeroth (Fig.~\ref{fig:device3_voltage}(a)) and the first lobe (Fig.~\ref{fig:device3_voltage}(b)). In device 3 the gate voltage separation between the charge transitions was significantly smaller; therefore, while the conductance is higher in the first lobe it was not possible to unambiguously compare the behaviors in the $e$-state and the $o$-state valleys.

\subsection{Current-bias measurements}

\begin{figure*}
	\includegraphics[width=0.8\textwidth]{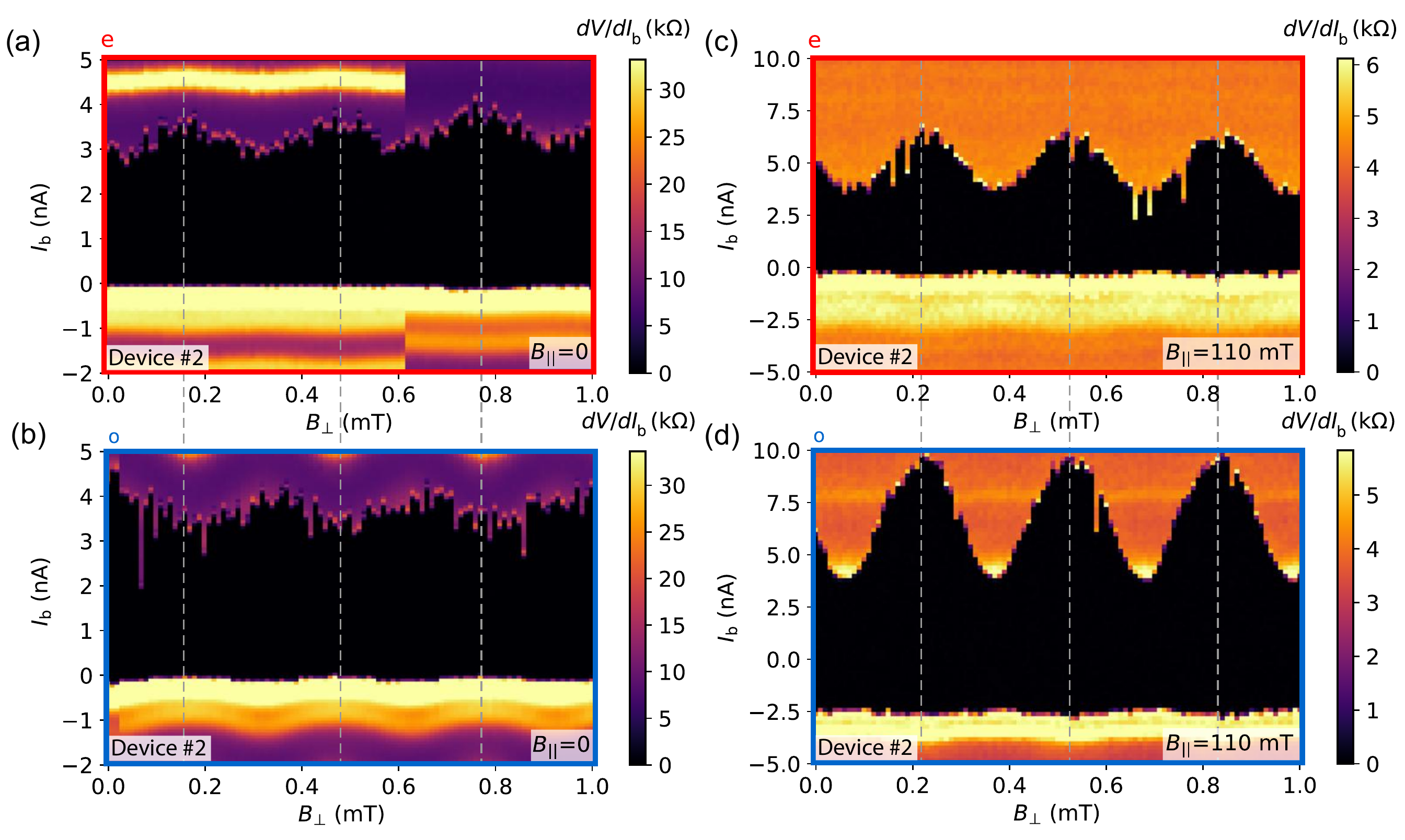}
	\caption{
		Differential resistance, $dV/dI_\mathrm{b}$, of device 2 as a function of bias current, $I_\mathrm{b}$, and perpendicular magnetic field, $B_\perp$, 
		(a,b) in the zeroth lobe, along cuts through the $e(o)$- state [red(blue) line in Fig.~\ref{fig:device2_voltage}(a)], showing relative $\pi$ phase shift,
		and (c,d) in the first lobe, along cuts through the $e(o)$-state [red(blue) dashed line in Fig.~\ref{fig:device2_voltage}(b)], showing absence of phase shift. Note that in the zeroth lobe ((a) and (b)) the magnitude of the oscillations in switching current are similar; however, in the first lobe the magnitude of the oscillations in the $o$-state (d) is significantly larger than in the $e$-state (c).
	}
	\label{fig:dev2_current}
\end{figure*}

\begin{figure*}[!ht]
	\includegraphics[width=\textwidth]{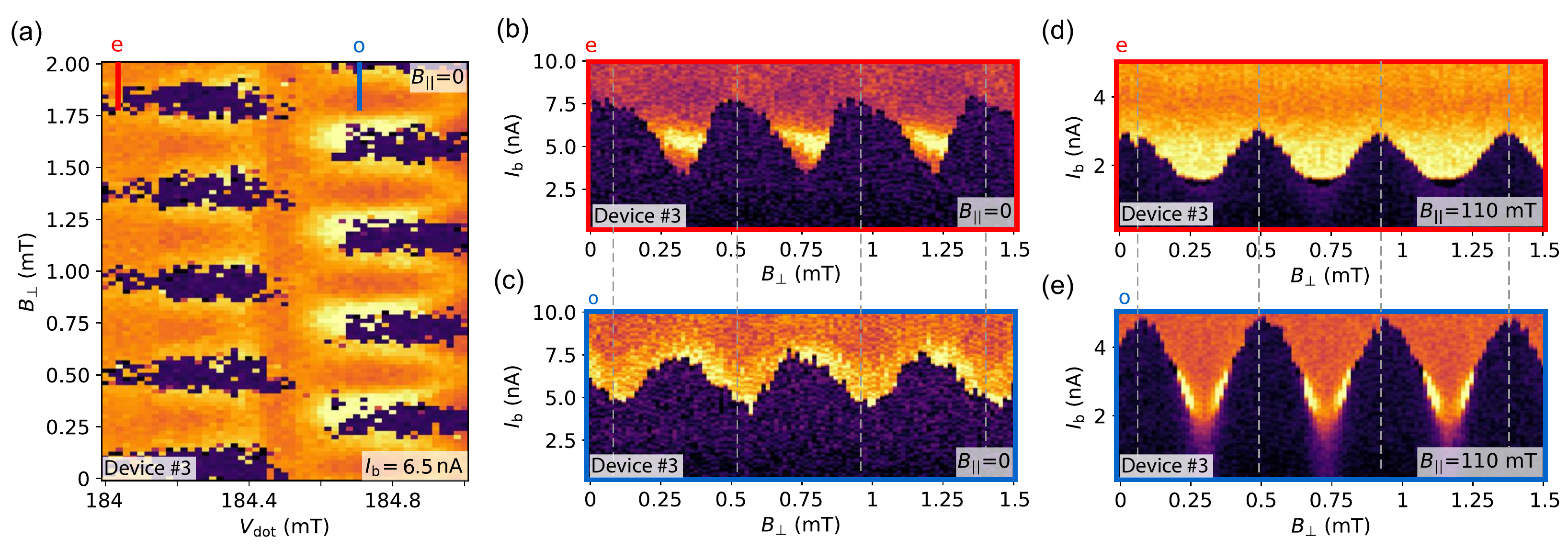}
	\caption{
		Differential resistance, $dV/dI_\mathrm{b}$, of device 3 as a function of bias current, $I_\mathrm{b}$, and perpendicular magnetic field, $B_\perp$.
		(a) $dV/dI_\mathrm{b}$ as a function of $\vdot$ and $B_\perp$ at fixed DC bias $I_\mathrm{b}=6.5\,$nA, the current is reset to zero before each pixel is acquired; a $\pi$ phase shift in the zero-resistance state is observed crossing from the $e$-state to the $o$-state. 
		(b,c) in the zeroth lobe, along cuts through the $e(o)$- state [red(blue) line in Fig.~\ref{fig:dev3_current}(a)], showing relative $\pi$ phase shift,
		and (d,e) in the first lobe, along cuts through the $e(o)$-state [red(blue) dashed line in Fig.~Fig.~\ref{fig:dev3_current}(a)], showing absence of phase shift. In the zeroth lobe ((b) and (c)) the magnitude of the oscillations in switching current are similar; however, in the first lobe the magnitude of the oscillations in the $o$-state (e) is significantly larger than in the $e$-state (d).
	}
	\label{fig:dev3_current}
\end{figure*} 

Current-bias spectroscopy measurements of the differential resistance, $dV/dI_{\mathrm b}$, for all devices was performed with the reference arm in a multichannel transmitting regime, controlled using $\vref$, so that $B_\perp$ primarily changes the phase across the dot-junction.

Figures~\ref{fig:dev2_current}(a,b) show current-bias measurements of the differential resistance, $dV/dI_{\mathrm b}$, for device 2 in the zeroth lobe as a function of $V_\mathrm{dot}$ and $B_\perp$, at the $\vdot$ positions indicated for the $e$ and $o$-states, respectively, in Fig.~\ref{fig:device2_voltage}(a). Similarly, Figs.~\ref{fig:dev2_current}(c,d) show $dV/dI_{\mathrm b}$ in the first lobe at the same gate voltages, in the $e$ and $o$-states. We review our conclusions based on these data below.

Figure~\ref{fig:dev3_current}(a) shows current-bias measurements of $dV/dI_{\mathrm b}$ for device 3 in the zeroth lobe as a function of $V_\mathrm{dot}$ and $B_\perp$, with a fixed DC current-bias of 5~nA, where the DC current-bias was reset to zero before each pixel was acquired. At the voltage where the charge transition crosses zero energy, we observed a $\pi$ shift of the supercurrent phase. This can be seen by comparing the phase of switching current oscillations between the $e$-state (Fig.~\ref{fig:dev3_current}(b)) and the $o$-state (Fig.~\ref{fig:dev3_current}(c)), at the $\vdot$ positions indicated in Fig.~\ref{fig:device3_voltage}. In the first lobe, the $\pi$ shift was absent as seen by comparing the $e$-state (Fig.~\ref{fig:dev3_current}(d)) and the $o$-state (Fig.~\ref{fig:dev3_current}(e)).

Reviewing the key observations relating to the switching current in device 1 we found that devices 2 and 3 reproduced the following behavior:
\begin{itemize}
	\item[(i)] In the zeroth lobe, a $\pi$ phase shift between the $e$-state and the $o$-state was observed; for device 2 compare Figs.~\ref{fig:dev2_current}(a,b), for device 3 compare Figs.~\ref{fig:dev3_current}(b,c).
	\item[(ii)] In the first lobe, no phase shift between the $e$-state and the $o$-state was observed; for device 2 compare Figs.~\ref{fig:dev2_current}(c,d), for device 3 compare Figs.~\ref{fig:dev3_current}(d,e).
	\item[(iii)] The absolute phases of critical current oscillations are aligned for both lobes and parities, with a $\pi$ phase shift for the $o$-state in the zeroth lobe; for device 2 see Fig.~\ref{fig:dev3_current}(b), and for device 3 see Fig.~\ref{fig:dev2_current}(b).
	\item[(iv)] The amplitude of the oscillatory component of the switching current is larger for the $o$-state than for the $e$-state; for device 2 compare Figs.~\ref{fig:dev2_current}(b,d), for device 3 compare Figs.~\ref{fig:dev3_current}(d,e).
\end{itemize}

\section{Additional Data for Device 1}

\begin{figure}[H]
	\centering
	\includegraphics[width=0.48\textwidth]{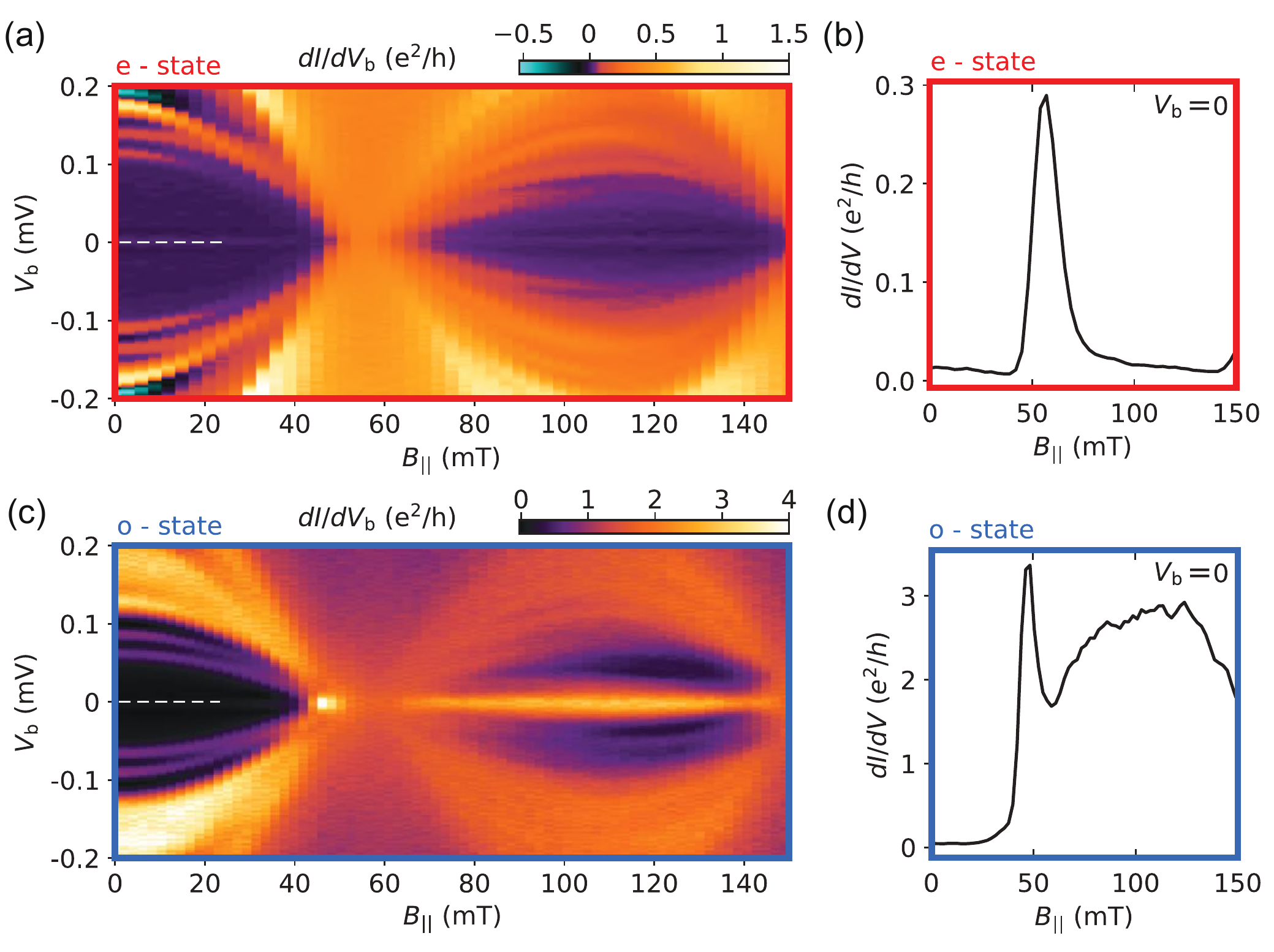}
	\caption{
		Device 1. Zero voltage bias cuts for even and odd states. (a) Differential resistance, $dV/dI_\mathrm{b}$, as a function of bias and axial magnetic field $B_{\parallel}$ in the even state. (b) Zero bias ($V_{\mathrm{b}}=0\,$) cut from (a) along the axial magnetic field $B_{\parallel}$. Same as (a) only in the odd state. (d) Same as (b) only in the odd state. Note bright feature at 46 mT and arc-shaped conductance in the first lobe, which roughly follows the size of the first-lobe gap.
	}
	\label{fig:dev1_zerobias_cuts}
\end{figure} 

\begin{figure}[H]
	\centering
	\includegraphics[width=0.5\textwidth]{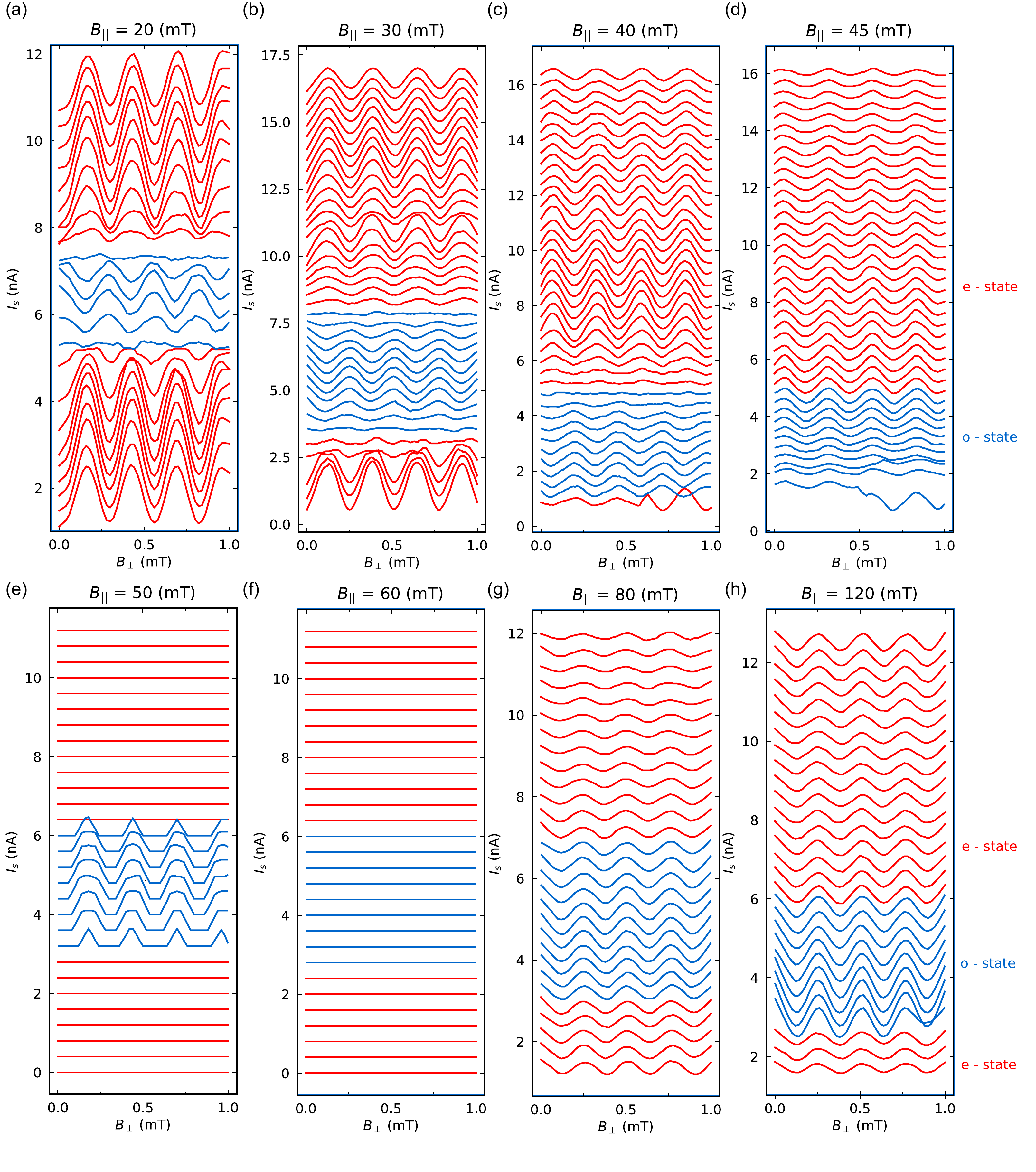}
	\caption{Device 1. Switching currents ($I_{s}$) across dot-junction charge occupancies (even-odd-even) with $\vdot$ at different axial magnetic fields $B_{\parallel}$. Each curve offset by 0.4 nA. (a) Switching currents at $B_{\parallel} = 20\,$mT crossing even (red), odd (blue) and back to even (red). A $\pi$-shift is visible between even and odd states. (b) Same as (a) only at $B_{\parallel} = 30\,$mT. (c) Same as (b) only at $B_{\parallel} = 40\,$mT. (d) Switching currents at $B_{\parallel} = 45\,$mT where the bright conductance peak was observed [see Fig. \ref{fig:dev1_zerobias_cuts}(d)], with no $\pi$-shift between even and odd charge occupancy. (e) Switching currents at $B_{\parallel} = 50\,$mT with finite current phase amplitude only in the odd state, with insets showing $B_{\perp}$ as a function of applied current bias. (f) Switching currents at $B_{\parallel} = 60\,$mT with no $B_{\perp}$ modulation of current phase amplitude. (g) Switching currents at $B_{\parallel} = 80\,$mT with no $\pi$ shift between the even and odd states. (h) Same as (g) only at $B_{\parallel} = 120\,$mT with no $\pi$ shift between the even and odd states. In addition, increase in current-phase amplitude modulation in the odd state relative to even state is observed.
	}
	\label{fig:dev1_CPR}
\end{figure} 

\begin{figure*}
	\includegraphics[width=1\textwidth]{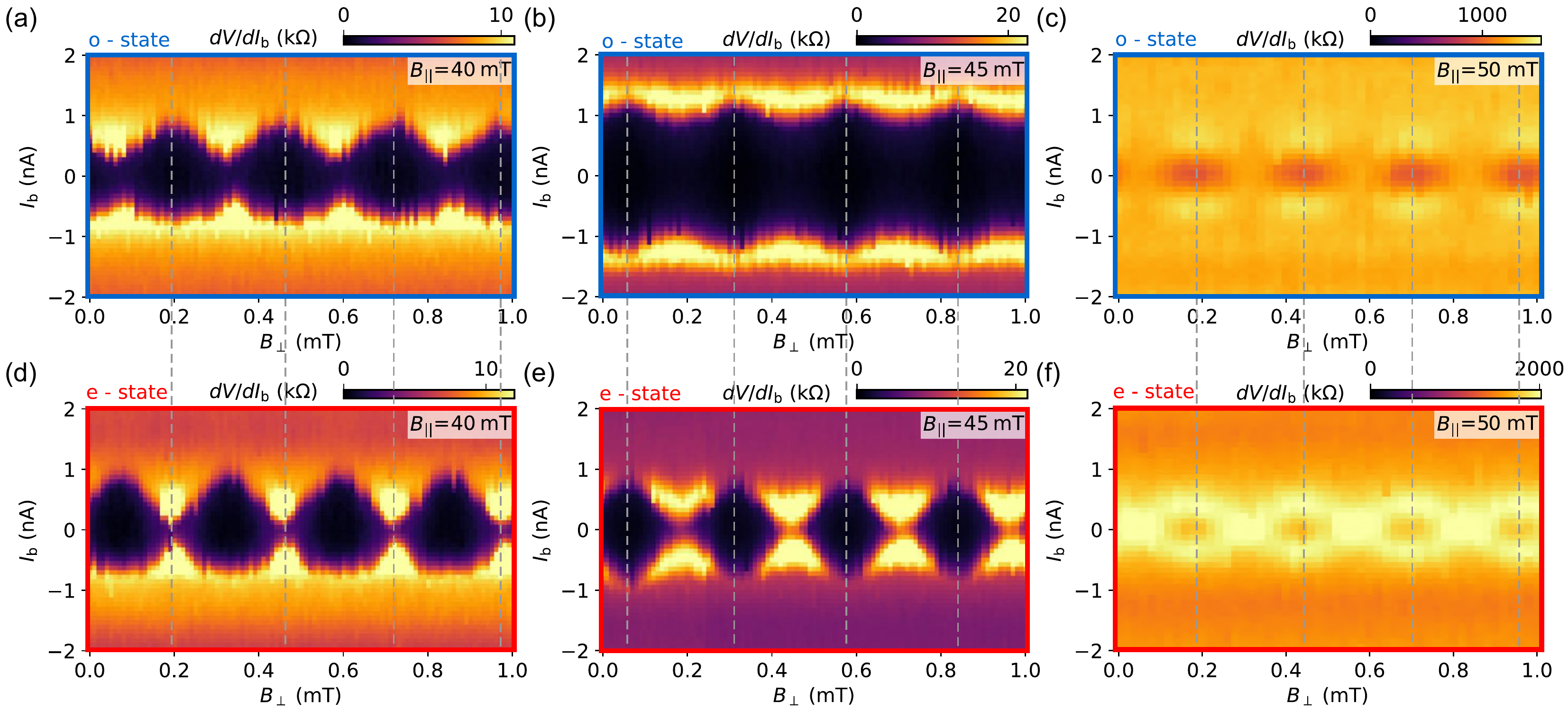}
	\caption{
		Device 1. Differential resistance, $dV/dI_{\mathrm b}$, as a function of current bias, $I_{\mathrm b}$, and perpendicular field $B_{\perp}$ at different fixed axial fields, $B_{\parallel}$. (a,d) Within the first lobe, $B_{\parallel}=40\,$mT, for even and odd dot-junction occupancies. Note $\pi$ phase shift between odd (a) and even (d) states. (b,e) On the bright spot,  $B_{\parallel}=45\,$mT, at the closing of the zeroth lobe [see Fig. \ref{fig:dev1_zerobias_cuts}(c-d)]. Note that even and odd states are in phase. (c, f) Entering the destructive regime, $B_{\parallel}=50\,$mT, with suppressed coherence effect, even and odd occupancies are in phase.
	}
	\label{fig:dev1_Phase_in_field}
\end{figure*}

\end{document}